\documentclass[aps,superscriptaddress,showpacs]{revtex4-1}
\usepackage{graphicx,epsfig,amssymb}
\usepackage{bm}
\usepackage{pstricks}
\usepackage{setspace}
\usepackage{graphicx}

\begin{document}
\title{Kondo Physics in a Rare Earth Ion with well localized {\it 4f } electrons}
\author{Jolanta Stankiewicz}
\email{jolanta@unizar.es}
\author{Marco Evangelisti}
\affiliation{Instituto de Ciencia de Materiales de Arag\'on and Departamento de F\'{i}sica de la Materia Condensada, CSIC--Universidad  de Zaragoza, 50009-Zaragoza, Spain}
\author{Zachary Fisk}
\affiliation{Department of Physics and Astronomy, University of California, Irvine, CA 92697, USA}
\author{Pedro Schlottmann}
\affiliation{Department of Physics, Florida State University, Tallahassee, FL 32306, USA}
\affiliation{National High Magnetic Field Laboratory, FL 32310, USA}
\author{Lev P. Gor'kov}
\affiliation{National High Magnetic Field Laboratory, FL 32310, USA} 
\affiliation{L. D. Landau Institute for Theoretical Physics of the RAS, Chernogolovka 142432, Russia} 

\date{\today}

\vspace{1cm}
\begin{abstract}

Dilute Nd in simple cubic LaB$_6$ shows electrical resistance and specific heat features at low temperature consistent with a Kondo scale of  $T_K \lesssim$ 0.3 K. Nd has a well localized {\it 4f }$^{ 3}$ {\it J} = 9/2 Hund's Rule configuration which is not anticipated to be Kondo coupled to the conduction electrons in LaB$_6$. We conjecture that the unexpected Kondo effect arises via participation of {\it 4f} quadrupolar degrees of freedom of the Nd crystal field ground state quartet.
\end{abstract}

\pacs{72.15.Qm, 75.20.Hr}
\maketitle

The Kondo effect underpins a broad range of correlated electron behavior,~\cite{pcol,bjon} including exotic superconductivity in {\it 4f} and {\it 5f} intermetallics,~\cite{hew} but is unexpected in materials with well-localized {\it f} levels. Spin interactions of {\it 4f}-electrons with conduction states are normally either of the Heisenberg exchange or of the Coqblin-Schrieffer type. The former exchange is ferromagnetic and arises from the Coulomb interaction in conjunction with the antisymmetry of the wave function. It scales with the so-called de Gennes factor, $(g-1)^2J(J+1)$ (where $g$ is the gyromagnetic ratio and $J$ is the total angular momentum of the state), and manifests itself in the Curie-Weiss magnetic susceptibility, spin disorder scattering in the electrical resistivity, and the depression of superconducting transition temperatures, for isostructural rare earths with well localized {\it 4f} levels. Ce and Yb do not follow this scaling in most cases (as of course divalent Eu does not as well), because their exchange is of the Coqblin-Schrieffer type,~\cite{coq} coming from the hybridization of the {\it 4f} levels with conduction electrons, leading to an antiferromagnetic exchange, the general condition for the development of the Kondo effect at low temperature. The {\it 4f} levels of Nd$^{3+}$ impurities in a metal are deep and only slightly hybridized.~\cite{muk} The exchange interaction is expected to be of the ferromagnetic Heisenberg type and, for this reason, the Kondo-like features for Nd impurities in LaB$_6$ we report in this Letter are surprising. What is essentially an old and well understood phenomenon makes a new appearance in LaB$_6$:Nd which is both unexpected and needing explanation.

The rare earth hexaborides crystallize in the CaB$_6$ simple cubic structure which can be pictured as a CsCl arrangement of B$_6$ octahedra and rare earth atoms. The boron network requires two electrons to form a closed shell configuration, leaving one conduction electron per unit cell in the hexaborides formed by trivalent rare earths. This electronic description has been well confirmed by band structure calculations~\cite{lang} as well as extensive de Haas-van Alphen measurements,~\cite{deursen,onuki,good} and agrees exactly with Hall effect measurements.~\cite{onuki,stan2} NdB$_6$ is a $T_N$ = 8.6 K type I antiferromagnet. The {\it 4f }$^{ 3}$ {\it J} = 9/2 crystal field scheme for the full cubic site symmetry of the Nd ion has been determined by both inelastic neutron~\cite{Loew} and Raman scattering~\cite{pofahl} experiments to be a ground state $\Gamma_8^{2}$ quartet, with the $\Gamma_8^{1}$ quartet and $\Gamma_{6}$  doublet lying at 135 K and 278 K above the ground state, respectively. An analysis of the temperature dependent specific heat using this magnetic level scheme, in addition to the lattice specific heat of LaB$_6$, led to the surprising conclusion that the electronic specific heat of NdB$_6$ above $T_N$ equaled 90 mJ/mole-Nd K$^2$.~\cite{stan1} This has prompted the investigation of the single ion properties of Nd dissolved in the isoelectronic, isostructural host LaB$_6$ reported here.  We find that Nd diluted in simple cubic LaB$_6$ exhibits electrical resistance, magnetic susceptibility, and specific heat features at low temperature, which are consistent with a Kondo scale of  $T_K \lesssim$ 0.3 K. We argue that quadrupolar interactions between the {\it 4f} and the conduction electrons, which give rise to the formation of the spin singlet, are  most likely responsible for this behavior.

All measurements were performed on single crystals grown from Al flux using high purity elemental rare earth from the Ames Laboratory.  Specific heat measurements using the relaxation method were carried out down to approximately 0.3~K in a commercial setup for the $0\leq \mu_{0}H\leq 9$~T magnetic field range. Magnetic susceptibility measurements were performed in a commercial superconducting quantum interference device magnetometer for temperatures down to 2~K, and in a homemade Hall microprobe magnetometer installed in a ${^3}$He setup for the lower temperature range. The four-probe resistivity was measured to low temperature in a ${^3}$He cryostat or in a PPMS system. Contact leads (25 $\mu$m gold wire) were soldered to the samples using pure indium. In our experiments, we used a dc (less than 1 mA) or low-frequency ac current. We etched the crystals in concentrated HCl to clean the surfaces before performing our measurements.

\begin{figure}
\includegraphics*[width=8.7cm]{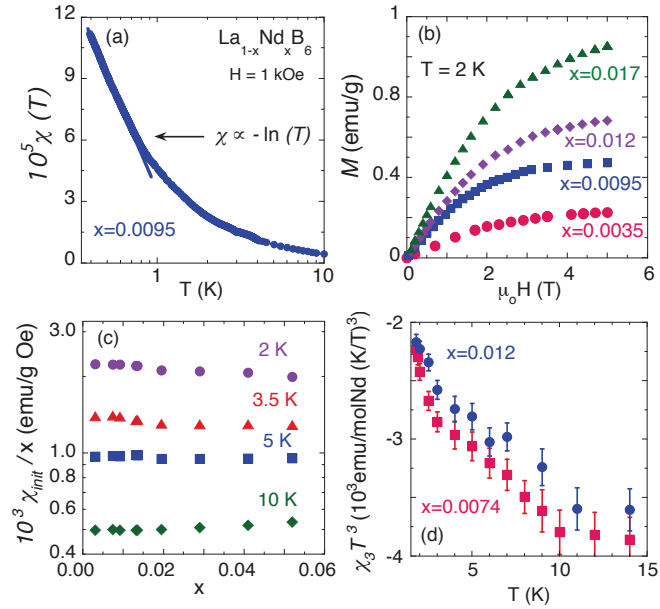}
\caption{(Color on-line)(a) - Low-temperature magnetic susceptibility for La$_{1-x}$Nd$_x$B$_6$ ($x=0.0095$) single crystal. A $\chi\propto -\ln(T)$ variation is shown for $T\lesssim$ 1~K. (b) - $M(H)$ at $T$ = 2~K for various La$_{1-x}$Nd$_x$B$_6$ single crystals. (c) - $\chi_{init} / $x versus x at various temperatures. (d) - The measured nonlinear susceptibility $\chi _{3}$ times $T^3$ at low temperatures.}
\label{M-T}
\end{figure}  

We measured electrical resistivity, magnetization and heat capacity on La$_{1-x}$Nd$_x$B$_6$  single crystals for $0\le x \lesssim 0.06$. In order to avoid any uncertainty in the Nd content, the very same samples were used in all experiments. The Nd compositions in the LaB$_6$ crystals were in general quite different from the nominal charge compositions. Therefore the Nd concentration was determined from low-temperature susceptibility and magnetization measurements, such as shown in Fig.~\ref{M-T}(a)-(b), using the crystal field scheme determined by neutron and Raman scattering in NdB$_6$. Since the effective moments for the $\Gamma_8^{2}$ and $\Gamma_{6}$ are quite close, nearly the same Nd composition would be determined from the susceptibility measurement assuming either ground state. The specific heat measurements confirmed unequivocally that the ground state of Nd ions in LaB$_6$ is the $\Gamma_8^{2}$ quartet as in pure NdB$_6$. Fig.~\ref{M-T}(c) shows that the initial susceptibility $\chi_{init}$ of our samples has no term proportional to $x^2$. Neither we find such term in the saturation magnetization which varies linearly with the Nd composition. We conclude therefore that the concentration of magnetic  pairs or clusters, which usually give rise to the  $x^2$ variation of magnetization, is negligible in our samples,~\cite{thol} in agreement with electron microprobe results. On the other hand, the induced magnetization in the paramagnetic state is $M = \chi_{1} H + \chi _{3}H^3+...$, where the nonlinear susceptibility $\chi _{3}$ probes quadrupolar fluctuations as shown for several rare-earth compounds.~\cite{morin,ram} The measured $\chi _{3}$ for La$_{1-x}$Nd$_x$B$_6$ single crystals is shown in Fig.~\ref{M-T}(d). Its behavior points unequivocally to the importance of the quadrupolar interactions as discussed below.

\begin{figure}
\includegraphics*[width=7cm]{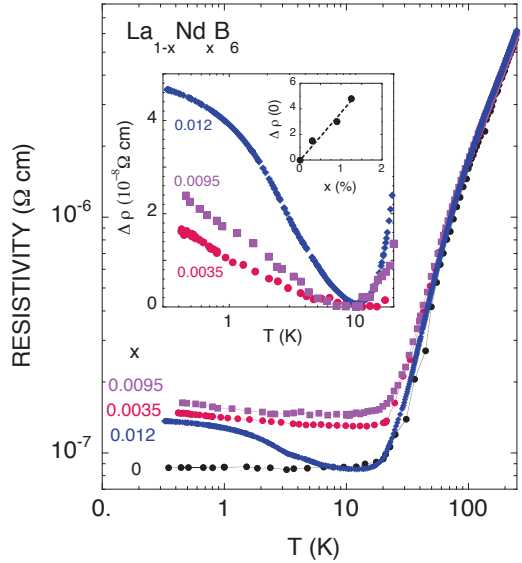}
\caption{(Color on-line) Electrical resistivity for La$_{1-x}$Nd$_x$B$_6$ single crystals. The main panel shows the temperature dependence of the electrical resistivity for La$_{1-x}$Nd$_x$B$_6$ single crystals in zero magnetic field. The Kondo contribution to the resistivity in the low-temperature region is shown in the inset. How this contribution varies with Nd content $x$ is plotted in the inset of the inset. }
\label{R-T}
\end{figure}
           
A Kondo rise in the low--temperature resistivity $\rho$ of dilute La$_{1-x}$Nd$_x$B$_6$ can be seen in Fig.~\ref{R-T}. The comparison with the LaB$_6$ resistivity data, also shown in Fig.~\ref{R-T}, rules out that the minimum in $\rho$ at approximately 10~K arises from impurities in the source material. Beyond this temperature, the resistivity of diluted alloys varies with $T$ as that of the pure LaB$_6$ sample. This indicates that the structural disorder is alike in all samples, and gives rise roughly to the same residual resistivity. The magnitude of the low--temperature rise in resistivity $\Delta$$ \rho$, plotted in the inset of Fig.~\ref{R-T}, is 3~$\mu$$\Omega$cm per mole Nd. For the well studied case of La$_{1-x}$Ce$_x$B$_6$ the value per Ce is 50 times larger,~\cite{winzer} where $T_K=0.86$~K.~\cite{schl} The excess resistivity goes nearly as $A-BT^{1/2}$ which points to non-Fermi liquid-like behavior. We observe the rise in the resistivity for a very small Nd--doping only $(x\lesssim 0.03)$ as shown in the inset of Fig.~\ref{R-T}.  With increasing $x$, one might expect sufficient strain in La$_{1-x}$Nd$_x$B$_6$ to lift the degeneracy of the ground-state quartet. If the four-fold degeneracy is central to the presence of the Kondo features found in these alloys, then the effect should strongly weaken with Nd concentration.
              
\begin{figure}
\includegraphics*[width=8.8cm]{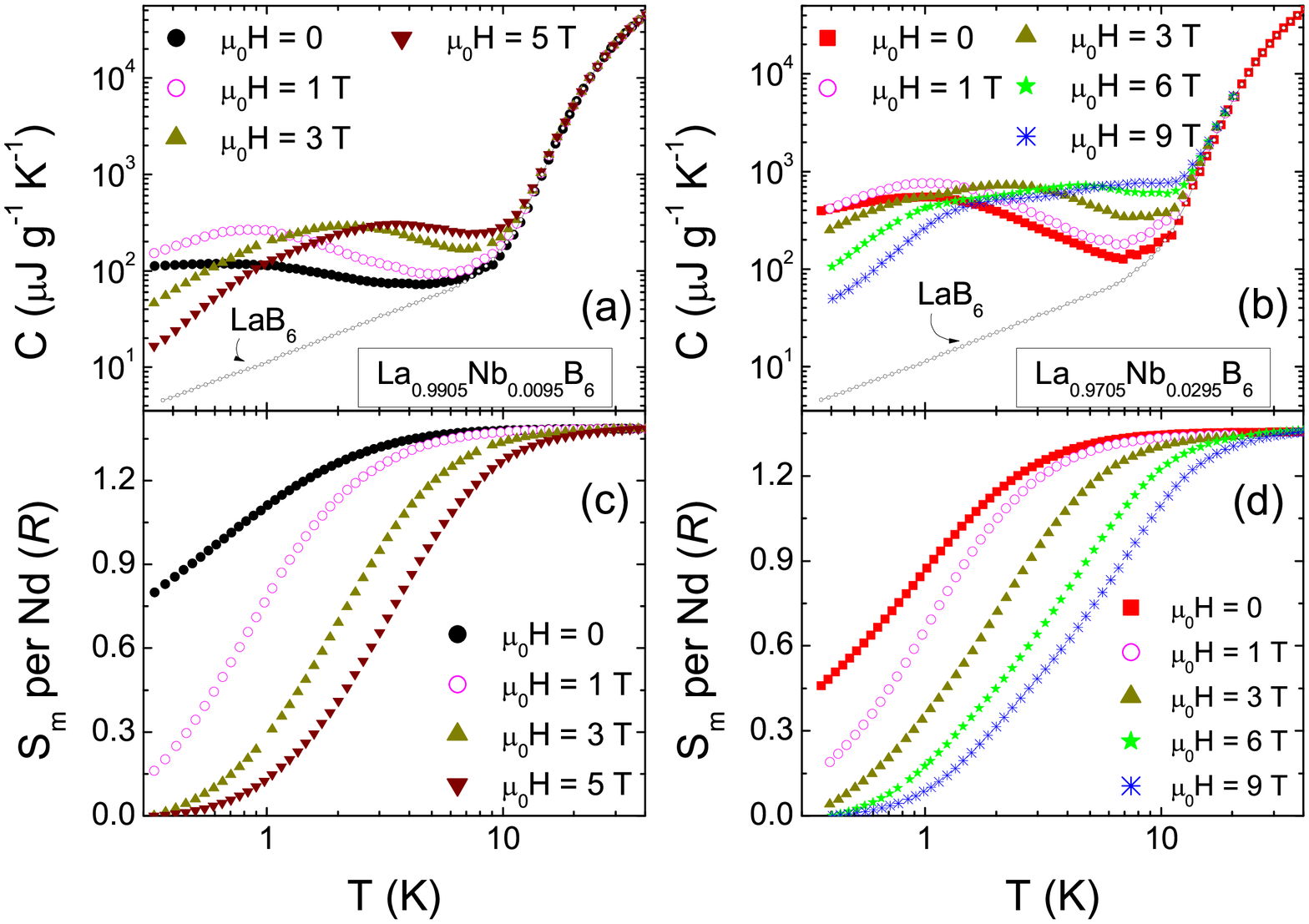}
\caption{(Color on-line)Specific heat for  La$_{1-x}$Nd$_x$B$_6$ single crystals. Top panels (a) and (b) show the temperature dependencies of the experimental specific heat $(C)$ for La$_{0.9905}$Nd$_{0.0095}$B$_6$ and La$_{0.9705}$Nd$_{0.0295}$B$_6$ single crystals, respectively, along with $C$ of the isomorphic LaB$_{6}$, at various magnetic fields, as labelled. Bottom panels (c) and (d) show how the excess (magnetic) entropy $S_{m}$ per Nd ion, normalized to the gas constant $R$, varies with temperature for the same single-crystals as in (a) and (b), respectively.}
\label{S-T}
\end{figure} 

In panels (a) and (b) of Fig.~\ref{S-T} we show the $T$-dependence of the specific heat $(C)$ of dilute La$_{1-x}$Nd$_x$B$_6$ for the $x=0.0095$ and 0.0295, respectively, in addition to isomorphic nonmagnetic LaB$_{6}$. The strong dependence of $C$ on $H$ at the lowest $T$ is brought about by the Nd ions. This magnetic contribution ($C_{m}$) superposes on the lattice contribution at higher temperatures, which is given by the data for LaB$_{6}$ compound. Furthermore, the electronic specific heat seen in LaB$_{6}$ (linear term $C_{el}\propto T$ in Fig.~3) should be expected to contribute likewise to La$_{1-x}$Nd$_x$B$_6$. Therefore, $C_{m}(T,H)$ is easily obtained by subtracting $C$ of LaB$_{6}$ from the total $C$ of La$_{1-x}$Nd$_x$B$_6$. Thus, by taking the integral $\int_{0}^{\infty}(C_{m}/T){\rm d}T$, the magnetic entropies ($S_{m}$) are calculated and depicted in panels (c) and (d) of Fig.~\ref{S-T} for $x=0.0095$ and 0.0295, respectively. For $\mu_{0}H> 1~{\rm T}$, $C_{m}(T,H)$ is well within our experimentally accessible temperature window. Indeed, extrapolating linearly $C_{m}/T$  as $T\rightarrow 0$~K  yields an $S_{m}$ content for $T\lesssim 0.3$~K that is less than 2\% of the total $S_m$ content. For $H=0$, the lack of experimental $C_m$ for $T\lesssim 0.3$~K has been taken into account by matching the limiting $S_m$ at high $T$ with the value obtained from the in-field data. 
After scaling $S_{m}$ to the proper Nd concentration, one can see that there is an entropy of $R~{\rm ln}(4)$ per mole Nd involved in the excess specific heat of La$_{1-x}$Nd$_x$B$_6$ above that of LaB$_6$, indicating that $(i)$ the  {\it 4f }$^{ 3}$ ground state is compensated by the conduction electrons and some other effects as $T\rightarrow 0$~K and $(ii)$ the ground state is a quartet, as is Ce in La$_{1-x}$Ce$_x$B$_6$.~\cite{gruhl} Assuming that about half of the zero-field entropy of a Kondo system is reached at the characteristic temperature $T_{K}$, the development of $S_{m}$ in panels (c) and (d) would suggest that $T_K$ is of order of $\lesssim 0.3$~K for the samples studied. 
    
 \begin{figure}
\includegraphics*[width=7.5cm]{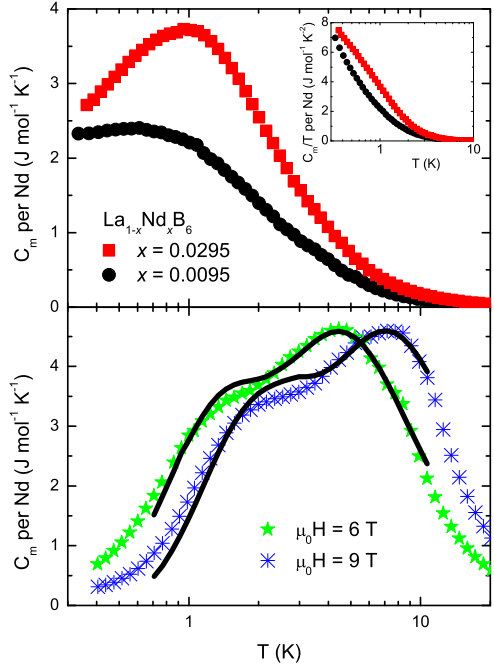}
\caption{(Color on-line)Magnetic specific heat for  La$_{1-x}$Nd$_x$B$_6$ single crystals. Top: Temperature dependencies of the experimental zero-field magnetic specific heat $(C_{m})$ per Nd ion for La$_{0.9905}$Nd$_{0.0095}$B$_6$ and La$_{0.9705}$Nd$_{0.0295}$B$_6$ single crystals, respectively. The same data, plotted as $C_{m}/T~vs.~{\log}(T)$, are in the inset. Bottom: Magnetic specific heat for La$_{0.9705}$Nd$_{0.0295}$B$_6$ single crystal between 0.3 and 20 K at various magnetic fields, as labelled. Solid lines are fits to the Schottky anomalies for the corresponding fields.}
\label{C-T}
\end{figure}

The zero-field magnetic contributions of Nd ions to the specific heat $(C_{m})$ are shown in the top panel of Fig.~\ref{C-T}. The dominant Schottky--type feature is seen to broaden and shift downward as $x$ decreases from $0.0295$ to 0.0095. By applying a magnetic field, these specific heat contributions gain height and get narrower, being no longer dependent on the Nd concentration for any field higher than $1-2$~T (Fig.~\ref{S-T}). For $\mu_{0}H\geq 6~{\rm T}$, we clearly distinguish two separate anomalies in $C_{m}$ (bottom panel of Fig.~\ref{C-T}), which we associate with the two doublets into which the $\Gamma_8^{2}$ quartet splits in a magnetic field. They have different {\it g}-factors, depending on the angle of the field with the crystallographic axis. The exact value of the {\it g}-factors for Nd also depends on the crystalline field splitting of the ion. For Nd, and for the field along (100) direction, the ratio of the {\it g}-factors is $2.94/1.92\approx 1.7$, using the Lea-Leask-Wolf table for $J=9/2$ and ${\rm x}=-0.90$.~\cite{lea} This is very close to what we obtain from a fitting of the Schottky four-level specific heat to the experimental LaB$_6$:Nd data at $\mu_{0}H=6$ and 9~T (solid lines in Fig.~\ref{C-T}).
	
The observed behavior agrees with the following picture.  Neglecting the possibility of a hybridization between the localized and conduction states, leaves us only with the ferromagnetic Coulomb exchange.  We ascribe the Kondo-like phenomena to the coupling of the conduction electrons with the quadrupolar momentum of the Nd-sites.  Because the magnetic degrees of freedom strongly couple with the quadrupolar ones, the degeneracy is lifted at $T=0$. A sufficiently high magnetic field and increasing $x$ break the Kondo-like coupling.  

The distance between Nd ions is not very large for the $x$ values considered and is a random variable. Hence, in zero-applied field, the Nd ions are not totally independent but see each other. It is therefore expected that the interaction between Nd ions leads to an internal field that locally splits the $\Gamma_8^{2}$ ground-state, according to a distribution whose width decreases with the increase of the Nd concentration. This interferes with the Kondo-like screening and leads to a distribution of ``Kondo temperatures'': $T_K$ differs at every site or locally in a region covering several unit cells. Both effects reduce the height of $C_{m}$ and make the ``Kondo resonance'' broader. Assuming that $(i)$ for $T<T_K$ the specific heat of an individual ion is proportional to $T/T_K$, and $(ii)$ the distribution of $T_K$ is reasonably flat, then $C_{m}/T\propto \ln(T_0/T)$, where $T_0$ in an average of $T_K$ (see inset of Fig.~\ref{C-T}).~\cite{kotliar} Similarly, the low-temperature zero-field susceptibility is proportional to $\ln(T_0/T)$ (see Fig.~\ref{M-T}(a)). A distribution of Kondo temperatures can also explain such behavior. The susceptibility for an individual ion for $T < T_K$ is proportional to $1/T_K$. A flat distribution of $T_K$ then again leads to $\chi \propto - \ln(T)$.\cite{kotliar} This agrees with our data. The excess resistivity $\Delta\rho\propto T^{1/2}$ over nearly a decade of temperature is also indicative of non-Fermi-liquid behavior. 

We should stress that the estimated value of $T_K$ ($ \lesssim 0.3$~K)  marks no more than a characteristic temperature scale (from competition between the ferromagnetic exchange and  scattering on the quadrupolar momentum of Nd-ions) and does not define the magnetic field scale same way as it does in the {\it s-d} model.  For LaB$_{6}$:Nd, the magnetic-field scale is given by the field that splits the $\Gamma_8^{2}$ quartet into two doublets. In Fig.~\ref {C-T}, the double-feature in the specific heat shows at $\mu_{0}H=6$~T, which corresponds to $T\approx 4$~K.

The low-$T$ magnetic susceptibility data follows a Curie-Weiss law down to 2~K with an  antiferromagnetic intercept of $\simeq -0.3$~K. The nonlinear susceptibility sheds additional light on the origin of the unexpected Kondo feature seen in the Nd alloys.  For noninteracting ions and within a quartet consisting of two doublets with Zeeman splitting, given by  $g_1$ and $g_2$, respectively, the linear and nonlinear susceptibilities are $\chi_{1}=((g_1^2+g_2^2)/8T)( \mu_B^2/k_B)$ and $\chi_{3} = -((g_1^4 + g_2^4 + 6 g_1^2 g_2^2)/(192 \times T^3))(\mu_B^4/k_B^3)$, respectively. The free-ion $\chi_{3}$ is always negative reflecting the negative curvature of the Brillouin function. A plot of $\chi_{3}T^3$ vs $T$ then gives a constant and deviations from this constant are consequently brought about by the Kondo-like screening. An inspection of Fig.~\ref{M-T}(d) shows
that the measured $\chi_{3}T^3$ for LaB$_6$:Nd is both negative and, for $T\lesssim 10$~K, monotonically increasing with decreasing temperature. The observed magnitude and temperature dependence of $\chi_{3}$ is similar for the other two samples studied. These results strongly favor the quadrupolar model in LaB$_6$:Nd.

Quadrupolar effects are strongly evident in the electronic properties of a number of rare earth intermetallics, most notably in the Pr skutterudites where they appear to give rise to heavy Fermion behaviour.~\cite{map,aoki} However, one does not find single-ion Kondo behavior in the properties of dilute Pr dissolved in the La skutterudite analogs.~\cite{rot} No heavy Fermion behavior has been seen in the stoichiometric Nd skutterudites. Raman experiments as well as detailed theoretical studies do indicate that one expects strong quadrupolar influence in the properties of NdB$_6$.~\cite{kubo,sera,pofahl} Note that for La$_{1-x}$Ce$_x$B$_6$ strong Kondo features exist for all Ce concentrations and for large $x$ antiferroquadrupolar order has been found.~\cite{Effantin}
     
 It is established that quadrupolar degrees of freedom play a fundamental role for Ce and Nd ions with $\Gamma_8$ ground-state; they manifest themselves in first place through interactions among the ions. Impurities strain the lattice and the local lattice distortions couple to the quadrupolar, but not the spin, degrees of freedom of the {\it 4f} electrons. The quadrupolar intersite interactions are probably the main reason for why it is hard to observe the single ion limit in both, LaB$_6$:Nd and LaB$_6$:Ce. The induced quadrupolar distortion could also lead to an interaction between the $\Gamma_8$ states and the conduction electrons. There is no consensus about the origin of the interactions. Kubo and Kuramoto~\cite{kubo} successfully described the excitation spectrum of NdB$_6$ using nearest-neighbour intersite exchange and quadrupolar interactions. A different approach emphasizing crystalline fields was proposed by Uimin and Brenig.~\cite{Uimin} For CeB$_6$, on the other hand, quadrupolar interactions between sites,~\cite{Uimin2} the RKKY interaction arising from the Coqblin-Schrieffer model,~\cite{Ohkawa,Schl} and a detailed group theoretical study~\cite{Shiina} have been presented. The well-localized nature of the {\it 4f} electrons in Nd compared to the considerable hybridization in Ce, suggests that the interaction mechanisms with the conduction states could be different for these two ions in the LaB$_6$ matrix.

We acknowledge support from grant MAT2008/03074, from the Ministerio de Ciencia e Innovaci\'{o}n  of Spain, and from National Science Foundation grant DMR-0801253. P. S. is supported by the US DoE under grant No. DE-FG02-98ER45707. NHMFL is supported by the NSF Cooperative Agreement No. DMR-0654118 and by the State of Florida. We are grateful to A. Cam\'on for his help in low-temperature magnetization measurements.

\end{document}